\documentclass{PoS}
\usepackage[alsoload=hep]{siunitx}

\DeclareSIUnit\eVperc{\eV\per\clight}
\DeclareSIUnit\tev{\tera\eV}
\DeclareSIUnit\gev{\giga\eV}
\DeclareSIUnit\mev{\mega\eV}
\DeclareSIUnit\gevperc{\giga\eV\per\clight}
\DeclareSIUnit\mevperc{\mega\eV\per\clight}
\DeclareSIUnit\gevpercsq{\giga\eV\per\clight\squared}
\DeclareSIUnit\gevpercsqsq{\giga\eV\squared\per\clight\tothe{4}}
\title{Latest results from the NA62 experiment at CERN}

\ShortTitle{Latest results from the NA62 experiment at CERN}

\author{\speaker{Nicolas Lurkin}\thanks{For the NA62 collaboration: 
R.~Aliberti, F.~Ambrosino, R.~Ammendola, B.~Angelucci, A.~Antonelli, G.~Anzivino, R.~Arcidiacono, T.~Bache, M.~Barbanera, J.~Bernhard
% ...
}\\
        School of Physics and Astronomy, University of Birmingham\\
        E-mail: \email{nicolas.lurkin@cern.ch}}

\abstract{NA62 is a fixed target kaon experiment at the CERN SPS which aims at measuring the branching ratio of the $K^+\to\pi^+\nu\bar{\nu}$ decay with \SI{10}{\%} precision. This ultra-rare kaon decay is theoretically extremely clean and an ideal place to look for physics beyond the Standard Model. The NA62 experiment has been running in 2016-2018 and has accumulated a large amount of kaon decays. The final result of the $K^+\to\pi^+\nu\bar{\nu}$ analysis using the 2016 dataset is discussed as well as prospects for the analysis of the 2017 dataset. NA62 also has sensitivity to various other rare and forbidden kaon decays, as well as some hidden sector models. The results of a search for the Lepton Number Violating decays $K^+\to\pi^-\ell^+\ell^+\ [\ell=e,\mu]$ is presented, along with new limits on the existence of a dark photon.
}

\FullConference{ALPS 2019 An Alpine LHC Physics Summit \\
		April 22 - 27, 2019\\
		Obergurgl, Austria}

\begin{document}

\section{Introduction}
The $K^+\to\pi^+\nu\bar{\nu}$ is a Flavour Changing Neutral Current process forbidden at tree level in the Standard Model (SM) and proceeding through penguin and box diagrams. It is highly suppressed due to the quadratic GIM mechanism and the small value of the CKM element $\left|V_{td}\right|$, which makes it an ideal channel to search for physics beyond the SM (BSM)\cite{RS,MSSM1,MSSM2,Zprime,lhiggs,LFUV}. The SM prediction is $\mathcal{B}(K^+\to\pi^+\nu\bar{\nu}) = \num{8.4(10)e-11}$\cite{buras}, where the dominant sources of uncertainties are the CKM parameters. The previous experimental result is $\mathcal{B}(K^+\to\pi^+\nu\bar{\nu}) = \left(17.3^{+11.5}_{-10.5}\right)\times10^{-11}$ obtained from 7 event candidates at the E787 and E949 experiments\cite{bnl}. The precision of the measurement does not allow us to draw any conclusion on any deviation from the SM prediction.

The total Lepton Number is a conserved quantity in the SM. However, Lepton Number Violation (LNV) is predicted in some BSM models. The kaon decays $K^+\to\pi^-\ell^+\ell^+\ [\ell=e,\mu]$ would then be allowed and proceed via an intermediate mediator such as a Majorana Neutrino\cite{shrock,atre}. Previous experimental results set the following limits at \SI{90}{\%} CL on the branching fraction of such channels: $\mathcal{B}(K^+\to\pi^-e^+e^+)<\num{6.4e-10}$\cite{bnl-lfu} and $\mathcal{B}(K^+\to\pi^-\mu^+\mu^+)<\num{8.6e-11}$\cite{na48-lfu}. 

An extension of the SM adding a new $U(1)$ gauge-symmetry sector is one of the proposed solutions to explain the abundance of dark matter in the universe. In the simplest realisation of this scenario\cite{u1,dp}, the vector mediator field $A'$ (dark photon) of mass $M_{A'}$ associated to the new $U(1)$ interacts with the SM through a kinetic mixing term whose strength is given by the parameter $\epsilon$. In such a case, it is possible for the $A'$ to be produced in the decay chain $K^+\to\pi^+\pi^0,\ \pi^0\to A'\gamma$ where
$\mathcal{B}(\pi^0\to A'\gamma) = 2\epsilon^2\left(1-\frac{M_{A'}^2}{M_{\pi^0}^2}\right)^3\times \mathcal{B}(\pi^0\to\gamma\gamma)\ .$
With the assumption that the $A'$ does not decay into SM particles inside the detector, a missing energy signature might reveal its presence.

\section{The NA62 experiment}
NA62 is a fixed target experiment at the CERN SPS whose aim is to measure $\mathcal{B}(K^+\to\pi^+\nu\bar{\nu})$ with \SI{10}{\%} precision. In order to collect the necessary $10^{13}$ decays, the experiment is served by the \SI{400}{\gevperc} proton beam of the SPS. The protons impinge on a beryllium target and produce an intense secondary hadron beam at \SI{75}{\gevperc} with \SI{1}{\%} momentum spread (RMS) and containing about \SI{6}{\%} of kaons.


The incoming kaon is tagged by a differential Cherenkov detector (KTAG) and its momentum is measured by the GigaTracker (GTK) spectrometer. Those detectors are followed by a \SI{110}{\m} long vacuum tank which contains the \SI{60}{\m} fiducial decay volume (FV). A STRAW spectrometer at the end of the FV allows the measurement of momenta of the charged decay products, and a Ring-Imaging Cherenkov detector (RICH) is used for their identification. A set of electromagnetic calorimeters (LAV, LKr, IRC, SAC) located around and at the end of the vacuum tube provide hermetic coverage for photons emitted in the $K^+$ decay with angles up to \SI{50}{\milli\radian}. Hadronic calorimeters and muon veto detectors (MUV1,2,3) are located after the LKr calorimeter to complement the particle identification system. Further counters (CHANTI, CHOD, MUV0, HASC) are installed at different locations for triggering purposes and to provide additional rejection power against beam interactions with the detector and multi-track kaon decays. A detailed description of the detector and the beam line can be found in \cite{na62_det}.

\section{Results from the $K^+\to\pi^+\nu\bar{\nu}$ analysis}
\subsection{Signal selection}
The signature of the signal consists of an incoming $K^+$, an outgoing $\pi^+$ and no additional activity in the detector. The detected incoming (upstream) track compatible with a kaon hypothesis and the outgoing (downstream) tracks associated in time are used to reconstruct the decay vertex at their closest distance of approach. The longitudinal position of the vertex
is required to be located within a \SI{50}{\m} long decay region. The identification of the downstream particle is performed by combining the RICH measurement with the output of a multivariate classifier using the LKr and the hadronic calorimeters, providing a $\pi^+$ identification efficiency of \SI{64}{\%} for a $\mu^+$ suppression of \num{1e8}. The remaining background consists mostly of the $K^+\to\pi^+\pi^0$ decay, which is suppressed by a factor of \num{3e8} thanks to the hermetic photon veto system (LAV, LKr, IRC, SAC) rejecting any in-time energy deposit. 

The main kinematic variable is the missing mass squared $m_\text{miss}^2 = (P_K-P_\pi)^2$. In order to protect against mis-reconstruction of the momenta, $m_\text{miss}^2$ is computed in three ways using different combination of $P_K$ and $P_\pi$ measurements. The kaon momentum $P_K$ can be measured directly with the GTK or taken as the nominal beam momentum measured using a sample of $K^+\to\pi^+\pi^+\pi^-$ decays. The pion momentum $P_\pi$ can be measured either with the STRAW or with the RICH. The analysis is performed in two separate regions in the $m_\text{miss}^2$ variables, on either side of the $K^+\to\pi^+\pi^0$ peak (R1 as \SIrange[open-bracket={[},close-bracket={]}]{0}{0.01}{\gevpercsqsq} and R2 as \SIrange[open-bracket={[},close-bracket={]}]{0.026}{0.068}{\gevpercsqsq}). These boundaries are driven by the missing mass squared resolution which is \SI{e-3}{\gevpercsqsq}. In both regions, the $\pi^+$ momentum must be in the range \SIrange[range-units=single]{15}{35}{\gevperc} to ensure at least \SI{40}{\gev} of electromagnetic energy in the calorimeters and to optimise the RICH $\pi^+/\mu^+$ separation.

\subsection{Results from the 2016 dataset}
The signal acceptance $A$, determined from Monte-Carlo (MC) simulations, is \SI{1}{\%} in R1 and \SI{3}{\%} in R2. The number of kaon decays in the FV is measured from a sample of $K^+\to\pi^+\pi^0$ decays selected with control triggers without the extra activity rejection and requiring $m_\text{miss}^2$ to be in the $K^+\to\pi^+\pi^0$ region. It is estimated to be $N_K = \num{1.21(4)e11}$. The single event sensitivity is $SES=1/(A\cdot\epsilon\cdot N_K) = \left(3.15\pm0.01_\text{stat}\pm0.24_\text{syst}\right)$, where $\epsilon$ takes into account the inefficiencies due to the trigger and the random veto caused by accidental activity. Multiplying this number by the SM branching fraction of the $K^+\to\pi^+\nu\bar{\nu}$ decay, the expected number of signal events is $0.267\pm0.001_\text{stat}\pm0.020_\text{sys}\pm0.032_\text{ext}$.

\begin{figure}
	\center
	\includegraphics[width=.55\linewidth]{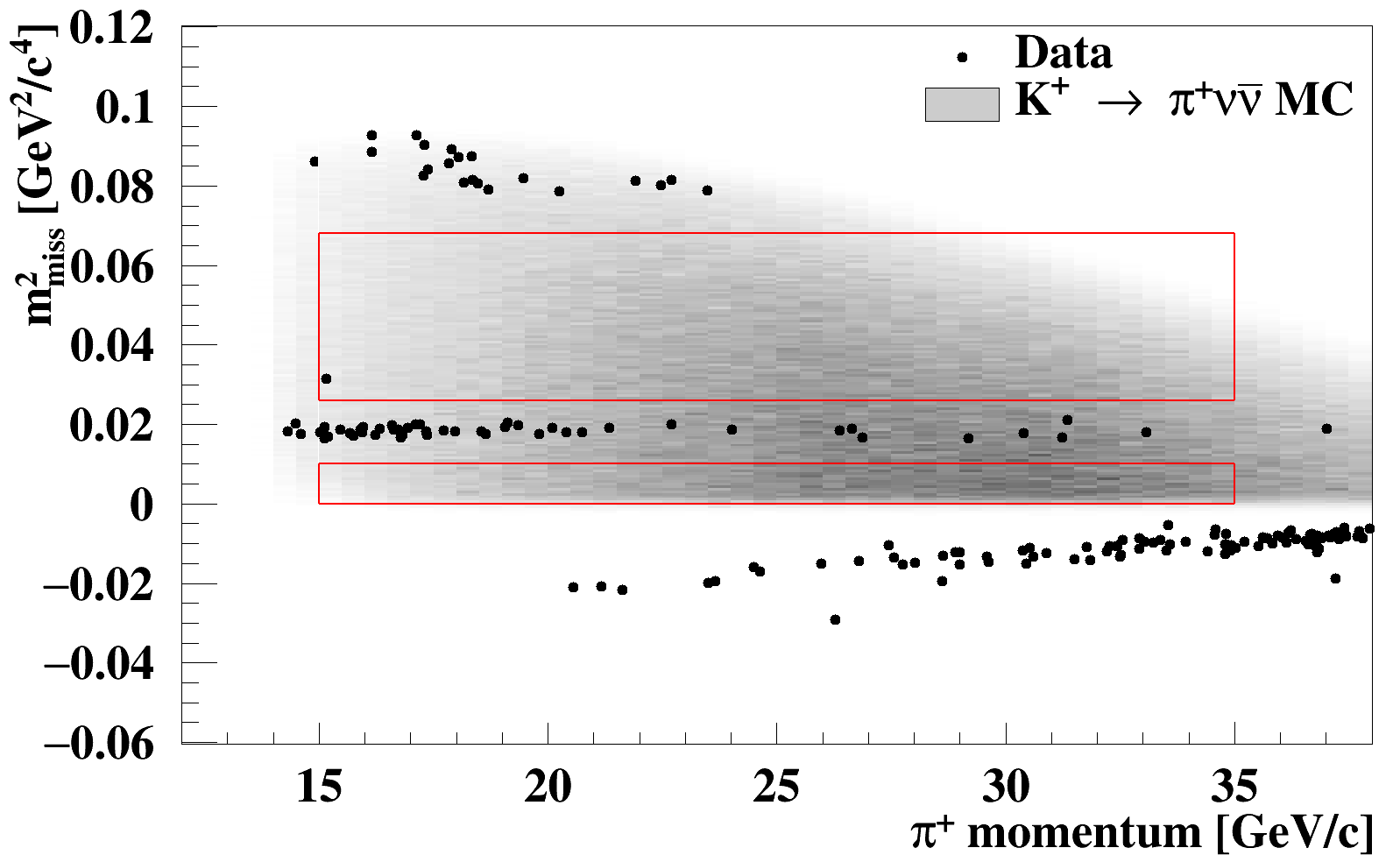}
	\caption{Reconstructed $m_\text{miss}^2$ as a function of the $\pi^+$ momentum for events satisfying the $K^+\to\pi^+\nu\bar{\nu}$ selection, except the $m_\text{miss}^2$ and $\pi^+$ momentum criteria. The grey area corresponds to the distribution of MC signal events. The red contours define the signal regions.}
	\label{pnn-result}
\end{figure}

The contribution of the $K^+\to\pi^+\pi^-e^+\nu$ decay to the background is evaluated from 600M MC decays, which show a good agreement across five validation samples. The other background estimates are obtained using data-driven techniques and validated on control regions around the signal regions. The upstream background is due to accidental activity in the detector and interactions of the beam particles in the upstream detectors. It is effectively suppressed by geometrical cuts and by the matching of the kaon and pion tracks. The background contributions are summarised in Table \ref{table-background}. One event was observed in R2, as shown in Figure \ref{pnn-result}, corresponding to a p-value of \SI{15}{\%} for the background only hypothesis and an observed upper limit of $\mathcal{B}(K^+\to\pi^+\nu\bar{\nu})<\num{14e-10}$ at \SI{95}{\%} CL \cite{pnn_paper}.

\begin{table}
	\center
	\begin{tabular}{l|l|l}
		\hline
		                                    &                                \multicolumn{2}{c}{Expected events (R1 + R2)}                                \\
		Process                             & \multicolumn{1}{c}{2016}                                   & \multicolumn{1}{c}{2017 (Preliminary)}         \\ \hline
		$K^+\to\pi^+\pi^0(\gamma)$          & $0.064\pm0.007_\text{stat}\pm0.006_\text{sys}$             & $0.35\pm0.02_\text{stat}\pm0.03_\text{sys}$    \\
		$K^+\to\mu^+\nu(\gamma)$            & $0.020\pm0.004_\text{stat}\pm0.006_\text{sys}$             & $0.16\pm0.01_\text{stat}\pm0.05_\text{sys}$    \\
		$K^+\to\pi^+\pi^+\pi^-$             & $0.002\pm0.001_\text{stat}\pm0.002_\text{sys}$             & $0.015\pm0.008_\text{stat}\pm0.015_\text{sys}$ \\
		$K^+\to\pi^+\pi^-e^+\nu$            & $0.013^{+0.017}_{-0.012}|_\text{stat}\pm0.009_\text{sys}$  & $0.22\pm{0.08}_\text{stat}$                    \\
		$K^+\to\pi^0\ell^+\nu (\ell=\mu,e)$ & $<0.001$                                                   & $0.012\pm0.012_{\text{sys}}$                   \\
		$K^+\to\pi^+\gamma\gamma$           & $<0.002$                                                   & $0.005\pm0.005_\text{sys}$                     \\
		Upstream background                 & $0.050^{+0.090}_{-0.030}|_\text{stat}$                     & Analysis ongoing                               \\ \hline
		Total background                    & $0.152^{+0.092}_{-0.033}|_\text{stat}\pm0.013_\text{syst}$ &                                                \\ \hline
	\end{tabular} 
	\caption{Summary of the background estimates summed over the two signal regions for the analysis of the 2016 dataset (2nd column) and preliminary estimates based on the 2017 dataset (3rd column).}
	\label{table-background}
\end{table}

\subsection{Future prospects}
The analysis of the 2017 data is ongoing with a 2016-like event selection. The analysis shows similar performance but benefits from several improvements in the pileup treatment for the IRC and SAC, in the LKr reconstruction and the usage of the RICH variables. Overall the $\pi^0$ rejection is \SI{40}{\%} more efficient. The background is currently under study and the preliminary estimates are summarised in Table \ref{table-background}. In particular, thanks to the range of beam intensities collected in 2017, investigations on intensity related effects are being made. They suggests that $\pi^+$ efficiencies, and $\pi^0$ and $\mu$ rejections have no or very little dependence on the beam intensity. In any case, both the signal and background scale linearly with the intensity. The data sample collected in 2017 is about 10 times bigger than the one used for the 2016 result. Considering $N_K = \num{1.3(1)e12}$ and $SES = \num{0.34(4)e-10}$, the expected number of SM events in 2017 is \num{2.5(4)}.


\section{Lepton Number Violation results}
\begin{figure}
	\center
	\includegraphics[width=.35\linewidth]{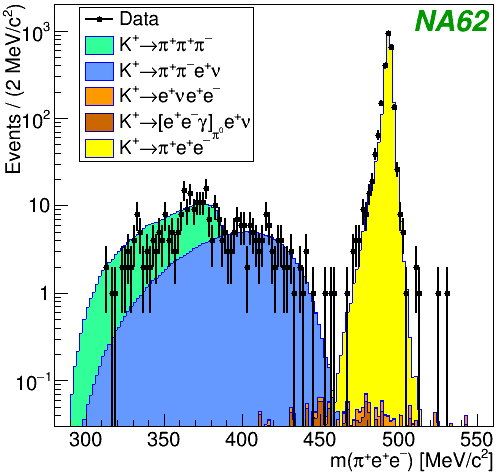}
	\includegraphics[width=.35\linewidth]{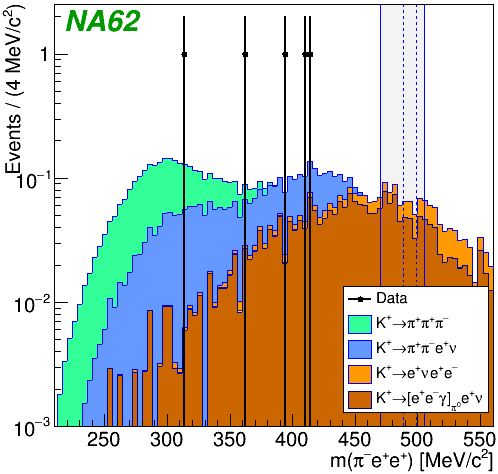}\\
	\includegraphics[width=.35\linewidth]{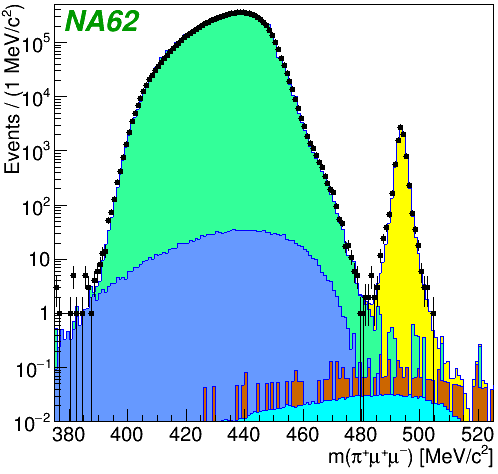}
	\includegraphics[width=.35\linewidth]{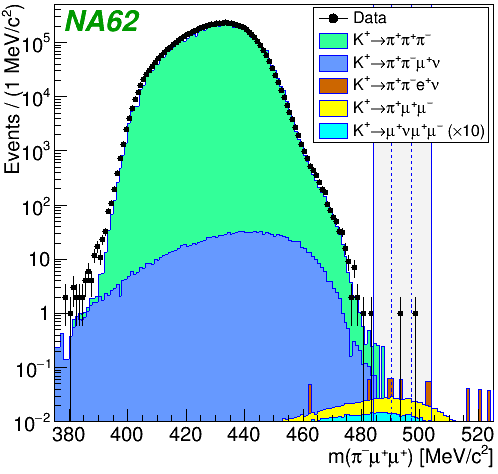}\\
	\caption{Reconstructed $\pi\ell\ell$ mass spectra for the LNV channels $K^+\to\pi^-e^+e^+$ (top right) and $K^+\to\pi^-\mu^+\mu^+$ (bottom right) with the partial 2017 data sample. The corresponding SM channels $K^+\to\pi^+e^+e^-$ (top left) and $K^+\to\pi^+\mu^+\mu^-$ (bottom left) are used for normalisation.}
	\label{lnv-result}
\end{figure}

Searches for the Lepton Number Violating (LNV) decay modes $K^+\to\pi^-\ell^+\ell^+ [\ell=e,\mu]$ were performed using a subset of the 2017 data sample corresponding to approximately 3 months of data taking. A blind analysis procedure was adopted and an event selection, similar for both channels, was developed. In both cases the corresponding $K^+\to\pi^+\ell^+\ell^-$ SM channel was used for normalisation, which results in the first order cancellation of systematic uncertainties. 
The main sources of background are $\pi^+$ mis-identification and decay in-flight. They are studied using a combination of MC and data-driven approaches. The reconstructed mass spectra are shown in Figure \ref{lnv-result}. 
In the $\ell=e$ case, the number of kaon decays $N_K=\num{2.14(7)e11}$, the signal acceptance of \SI{4.98}{\%} and the expected number of background events of \num{0.16(3)} for 0 observed signal events result in an upper limit on the branching fraction $\mathcal{B}(K^+\to\pi^-e^+e^+)<\num{2.2e-10}$ at \SI{90}{\%} CL. 
In the $\ell=\mu$ case, the number of kaon decays $N_K=\num{7.94(23)e11}$, the signal acceptance of \SI{9.81}{\%} and the expected number of background events of \num{0.91(41)} for 1 observed signal event result in an upper limit on the branching fraction $\mathcal{B}(K^+\to\pi^-\mu^+\mu^+)<\num{4.2e-11}$ at \SI{90}{\%} CL\cite{lfv-result}.

\section{Dark photon searches}
A sample of tagged $\pi^0$ from $K^+\to\pi^+\pi^0$ decays with exactly one $\gamma$ detected in the LKr is selected in a subset of the data collected by NA62 in 2016, corresponding to \SI{1}{\%} of the total collected in 2016-18. This sample is used to search for $A'$ peaks in the missing mass squared distribution $m_\text{miss}^2 = (P_K-P_\pi-P_\gamma)^2$, where $P_K, P_\pi$ and $P_\gamma$ are the 4-momenta of the beam kaon, the $\pi^+$ and the detected $\gamma$, respectively. A sliding search window of width $\pm1\sigma_{m_\text{miss}^2}$ is used to count the number of events for each $M_{A'}$ hypothesis, where the mass resolution is estimated with MC simulations. An independent sample of fully reconstructed $K^+\to\pi^+\pi^0$ is used for the normalisation.

The most abundant source of background is $\pi^0\to\gamma\gamma$ with one $\gamma$ undetected. A sample of $\pi^0\to\gamma\gamma$ decays where one $\gamma$ converts into $e^+e^-$ upstream of the LKr is selected. It is scaled to the signal sample in a side-band and used to estimate the background contribution in the search region. The side band extends between \SIrange[range-phrase=\ and\ ,range-units=single]{5e-5}{75e-5}{\gevpercsqsq}, while the search region extends between \SIrange[range-phrase=\ and\ ,range-units=single]{75e-5}{1765e-5}{\gevpercsqsq}. A CLs statistical treatment is applied and no statistically significant excess is observed, allowing the setting of new upper limits at \SI{90}{\%} CL on $\epsilon^2$ at the level of \numrange{e-7}{e-6} over the dark photon mass range \SIrange{60}{110}{\mevperc\squared}\cite{dp-result}.


\section{Conclusion}
The final result of the $K^+\to\pi^+\nu\bar{\nu}$ analysis from the 2016 dataset was reported. One candidate signal event was found, resulting in an upper limit on $\mathcal{B}(K^+\to\pi^+\nu\bar{\nu})$ of \num{14e-10} at \SI{95}{\%} CL. Preliminary estimates of the background level in the 2017 data were shown. The final results of searches for LNV kaon decay modes were presented, where the following upper limits at \SI{90}{\%} CL are set: $\mathcal{B}(K^+\to\pi^-e^+e^+)<\num{2.2e-10}$ and $\mathcal{B}(K^+\to\pi^-\mu^+\mu^+)<\num{4.2e-11}$. Finally limits on the $\epsilon^2$ kinetic mixing parameter at the level of \numrange{e-7}{e-6} are reported for dark photon masses in the range \SIrange{60}{110}{\mevperc\squared}.


\begin{thebibliography}{99}
\bibitem{RS}
M. Blanke \textit{et al.},
\emph{JHEP} {\bf 0903} (2009) 108
[{\tt hep-ph/0812.3803}]

\bibitem{MSSM1}
G. Isidori \textit{et al.},
\emph{JHEP} {\bf 0608} (2006) 064
[{\tt hep-ph/060474}]

\bibitem{MSSM2}
T. Bla\v{z}ek \textit{et al.},
\emph{Int.J.Mod.Phys} {\bf A29} (2014) no.27, 1450162
[{\tt hep-ph/1410.0055}]

\bibitem{Zprime}
A. Buras \textit{et al.},
\emph{JHEP} {\bf 1511} (2015) 166
[{\tt hep-ph/1507.08672}]

\bibitem{lhiggs}
M. Blanke \textit{et al.},
\emph{Eur.Phys.J.} {\bf C76} (2016) no.4, 182
[{\tt hep-ph/1507.06316}]

\bibitem{LFUV}
M. Bordone \textit{et al.},
\emph{Eur.Phys.J.} {\bf C77} (2017) no.9, 618
[{\tt hep-ph/1705.10729}]

\bibitem{buras}
A. Buras \textit{et al.},
\emph{JHEP} {\bf 1511} (2015) 033
[{\tt hep-ph/1503.02693}]

\bibitem{bnl}
A.V. Artamonov \textit{et al.},
\emph{Phys.Rev.D} {\bf 79} (2009) 092004
[{\tt hep-ex/0903.0030}]

\bibitem{shrock}
L. S. Littenberg \textit{et al.},
\emph{Phys.Lett. B} {\bf 491} (2000) 285
[{\tt hep-ph/0005285}]

\bibitem{atre}
A. Atre \textit{et al.},
\emph{JHEP} {\bf 0905} (2009) 030
[{\tt hep-ph/0901.3589}]

\bibitem{bnl-lfu}
R. Appel \textit{et al.},
\emph{Phys.Rev.Lett.} {\bf 85} (2000) 2877.
[{\tt hep-ex/0006003}]

\bibitem{na48-lfu}
J.R. Batley \textit{et al.}
\emph{Phys.Lett. B} {\bf 769} (2017) 67
[{\tt hep-ex/1612.04723}]

\bibitem{u1}
B. Holdom,
\emph{Phys.Lett. B} {\bf 166} (1986) 196

\bibitem{dp}
B. Batell \textit{et al.},
\emph{Phys.Rev. D} {\bf 80} (2009) 095024
[{\tt hep-ph/0906.5614}]

\bibitem{na62_det}
E. Cortina Gil \textit{et al} 
\emph{JINST} {\bf 12} (2017) no.05, P05025
[{\tt physics.ins-det/1703.08501}]

\bibitem{pnn_paper}
E. Cortina Gil \textit{et al} 
\emph{Phys.Lett.} {\bf B791} (2019) 156-166
[{\tt hep-ex/1811.08508}]

\bibitem{lfv-result}
E. Cortina Gil \textit{et al} 
[{\tt hep-ex/1905.07770}]

\bibitem{dp-result}
E. Cortina Gil \textit{et al} 
\emph{JHEP} {\bf 1905} (2019) 182
[{\tt hep-ex/1903.08767}]


\end{thebibliography}
\end{document}